\documentclass[12pt]{article}
\usepackage{amssymb,amsmath,epsfig}
\allowdisplaybreaks
\begin{document}
\title{\bf On the Stability of Einstein Universe in $f(R,T,R_{\mu\nu}T^{\mu\nu})$ Gravity}

\author{M. Sharif \thanks{msharif.math@pu.edu.pk} and Arfa Waseem
\thanks{arfawaseem.pu@gmail.com}\\
Department of Mathematics, University of the Punjab,\\
Quaid-e-Azam Campus, Lahore-54590, Pakistan.}
\date{}

\maketitle

\begin{abstract}
This paper investigates the existence and stability of Einstein
universe in the context of $f(R,T,Q)$ gravity, where
$Q=R_{\mu\nu}T^{\mu\nu}$. Considering linear homogeneous
perturbations around scale factor and energy density, we formulate
static as well as perturbed field equations. We parameterize the
stability regions corresponding to conserved as well as
non-conserved energy-momentum tensor using linear equation of state
parameter for particular models of this gravity. The graphical
analysis concludes that for a suitable choice of parameters, the
stable regions of the Einstein universe are obtained.
\end{abstract}

\section{Introduction}

The fact that our universe is going through an accelerated expansion
is one of the most spectacular discoveries of modern cosmology. This
motivated many researchers to investigate the reason behind the
phase of accelerated expansion. It is claimed that dark energy (DE)
having large negative pressure with repulsive effects is responsible
for the current expanding behavior of the universe. In order to
explore the mysterious nature of DE, several proposals have been
studied including modified theories of gravity as the inspiring
approach. These modified theories are the generalizations of the
Einstein-Hilbert action like $f(R)$ ($R$ indicates Ricci scalar)
\cite{1}, $f(\mathcal G)$ ($\mathcal G$ represents Gauss-Bonnet
invariant term) \cite{2}, $f(\mathcal T)$ ($\mathcal T$ shows
torsion) \cite{3}, $f(R,T)$ ($T$ describes the trace of
energy-momentum tensor (EMT)) \cite{4}, $f(\mathcal G,T)$ \cite{5}
and $f(R,T,Q)$ ($Q$ is the contraction of Ricci tensor and EMT)
gravity \cite{6}.

The $f(R,T,Q)$ gravity, an extension of $f(R,T)$ gravity, has gained
much attention due to its strong non-minimal coupling between
gravity and matter fields. Odintsov and
S$\acute{a}$ez-G$\acute{o}$mes \cite{7} explored matter instability,
$\Lambda$CDM model and de Sitter solutions in this modified theory.
Sharif and Zubair investigated the validity of thermodynamical laws
\cite{8} and derived the energy conditions \cite{9} for two
different models of this gravity. The isotropic as well as
anisotropic physical behavior of compact relativistic objects are
also discussed in this gravity \cite{10}. Baffou et al. \cite{11}
explored the stability analysis of this modified theory
corresponding to de Sitter and power-law solutions using
perturbation approach and found stable solutions. Recently, Yousaf
et al. \cite{12} studied the stability of cylindrical system for a
particular $f(R,T,Q)$ model and examined the instability constraints
at Newtonian and post-Newtonian limits.

The emergent universe scenario (which helps to resolve the issue of
big-bang singularity) is based on the existence as well as stability
of the Einstein universe (EU) against all kinds of perturbations. In
general relativity (GR), the idea of this emergent universe is not
proved successful due to unstable EU against homogenous
perturbations. The existence and stability of EU in modified
theories is of great importance. The stability of EU has been
checked in brane-world models, GR with small inhomogeneous vector
and tensor perturbations, GR with variable pressure, Einstein-Cartan
theory, loop quantum cosmology etc \cite{13}. B\"{o}hmer et al.
\cite{14} examined the stability of EU against linear perturbations
in $f(R)$ theory and found the existence of stable regions for
particular models of this theory. Goswami et al. \cite{15} observed
the existence and stable modes of EU in the context of fourth-order
modified theory. Goheer and his collaborators \cite{16} found stable
solutions of EU corresponding to power-law model in $f(R)$ gravity.
B\"{o}hmer and Lobo \cite{17} discussed existence as well as
stability of EU in $f(\mathcal G)$ gravity and obtained stable
states for different values of equation of state (EoS) parameter.

Carneiro and Tavakol \cite{18} investigated the existence of EU and
its stability under the effects of vacuum energy corresponding to
conformally invariant fields. Seahra and B\"{o}hmer \cite{19} showed
that stable EU solutions exist in $f(R)$ models only for perfect
fluid with linear EoS whereas they remain unstable against
inhomogeneous perturbations. Canonico and Parisi \cite{20} studied
the stability of EU in Ho\v{r}ava-Lifshitz gravity under some
certain conditions and similar analysis is performed in the
framework of massive gravity \cite{21}. B\"{o}hmer et al. \cite{22}
examined the stability of EU in hybrid metric-Palatini gravity using
homogeneous as well as inhomogeneous linear perturbations and found
the existence of large class of stable solutions. Li et al.
\cite{23} analyzed the stable modes for open as well as closed
universe by considering homogeneous perturbations in teleparallel
modified theory.

Huang et al. \cite{24} found stable EU solutions against anisotropic
and homogenous perturbations in the background of Jordan-Brans-Dicke
theory. The same authors \cite{25} also established the unstable
solutions for open universe and stable regions for closed universe
in Gauss-Bonnet gravity. B\"{o}hmer et al. \cite{26} examined the
stability modes against homogeneous as well as inhomogeneous
perturbations in scalar-fluid theories and obtained stable and
unstable results corresponding to inhomogeneous and homogeneous
perturbations, respectively. Darabi and his collaborators \cite{27}
studied the existence of EU and its stability in the framework of
Lyra geometry using scalar, vector and tensor perturbations with
suitable choice of parameters. Shabani and Ziaie \cite{28} discussed
the stable solutions of EU in $f(R,T)$ gravity which were unstable
in $f(R)$ gravity. Sharif and Ikram \cite{a} investigated the
stability of EU against linear homogeneous perturbations in
$f(\mathcal G,T)$ gravity. They found that stable EU solutions exist
and their results reduce to $f(\mathcal G)$ gravity in the absence
of matter-curvature coupling.

In this paper, we explore the stability of EU by applying
homogeneous linear perturbations in the framework of $f(R,T,Q)$
gravity. This study would help to investigate the effects of strong
non-minimal coupling of matter and geometry on the stability of EU.
The format of this paper is as follows. In the next section, we
formulate the corresponding field equations of this theory. Section
\textbf{3} deals with the stability of EU for both conserved and
non-conserved EMT. In the last section, we summarize our concluding
remarks.

\section{Formalism of $f(R,T,Q)$ Gravity}

The action for $f(R,T,Q)$ gravity is defined as \cite{6}
\begin{equation}\label{1}
S=\frac{1}{2\kappa^{2}}\int d^4x \sqrt{-g}[ f(R,T,Q)+
\mathcal{L}_{m}],
\end{equation}
where $\kappa^{2}$ and $\mathcal{L}_{m}$ represent coupling constant
and matter Lagrangian density, respectively. The EMT corresponding
to $\mathcal{L}_{m}$ is given by \cite{29}
\begin{equation}\label{2}
T^{\mu\nu}=\frac{2}{\sqrt{-g}}\frac{\delta(\sqrt{-g}\mathcal{L}_m)}{\delta
g_{\mu\nu}}=g^{\mu\nu}\mathcal{L}_{m}
+\frac{2\delta\mathcal{L}_{m}}{\delta g_{\mu\nu}}.
\end{equation}
Varying the action (\ref{1}) with respect to $g_{\mu\nu}$, we obtain
the field equations
\begin{equation}\label{3}
G_{\mu\nu}= R_{\mu\nu}-\frac{1}{2}Rg_{\mu\nu}=T_{\mu\nu}^{eff},
\end{equation}
where the effective EMT is of the form
\begin{eqnarray}\nonumber
T_{\mu\nu}^{eff}&=&\frac{1}{f_{R}-f_{Q}\mathcal{L}_{m}}\left[(1+f_{T}
+\frac{1}{2}Rf_{Q})T_{\mu\nu}+\left\{\frac{1}{2}(f-Rf_{R})-\mathcal{L}_{m}
f_{T}\right.\right.\\\nonumber&-&\left.\left.\frac{1}{2}\nabla_{\alpha}
\nabla_{\beta}(f_{Q}T^{\alpha\beta})\right\}g_{\mu\nu}-(g_{\mu\nu}
\Box -\nabla_{\mu}\nabla_{\nu})f_{R}+\nabla_{\alpha}\nabla_{(\mu}
[T_{\nu)}^{\alpha}f_{Q}]\right.\\\label{4}&-&\left.\frac{1}{2}
\Box(f_{Q}T_{\mu\nu})-2f_{Q}R_{\alpha(\mu}T_{\nu)}^{\alpha}
+2(f_{T}g^{\alpha\beta}+f_{Q}R^{\alpha\beta})\frac{\partial^{2}
\mathcal{L}_{m}}{\partial g^{\mu\nu}\partial
g^{\alpha\beta}}\right].
\end{eqnarray}
The subscripts of generic function $f$ show derivative with respect
to $R,~Q$ and $T$. The covariant divergence of field equation
(\ref{4}) is given by
\begin{eqnarray}\nonumber
\nabla^{\mu}T_{\mu\nu}^{eff}&=&\frac{2}{2(1+f_{T})+Rf_{Q}}\left[\nabla_{\mu}
(f_{Q}R^{\sigma\mu}T_{\sigma\nu})+\nabla_{\nu}(\mathcal{L}_{m}f_{T})
-G_{\mu\nu}\nabla_{\mu}(f_{Q}\mathcal{L}_{m})\right.\\\label{5}&-&
\left.\frac{1}{2}(f_{Q}R_{\alpha\beta}+f_{T}g_{\alpha\beta})
\nabla_{\nu} T^{\alpha\beta}-\frac{1}{2}[\nabla^{\mu}(Rf_{Q})
+2\nabla^{\mu}f_{T}]T_{\mu\nu}\right].
\end{eqnarray}

The line element for closed FRW universe model is
\begin{equation}\label{6}
ds^{2}=dt^{2}-a^{2}(t)\left(\frac{1}{1-r^{2}}dr^{2}
+r^{2}(d\theta^{2}+\sin^{2}\theta d\phi^{2})\right),
\end{equation}
where $a(t)$ represents the scale factor. The EMT for perfect fluid
is
\begin{equation}\label{7}
T_{\mu\nu}=(\rho + p)u_{\mu}u_{\nu}- pg_{\mu\nu},
\end{equation}
where $\rho,~p$ and $u_{\mu}$ indicate energy density, pressure and
four velocity, respectively. As we are interested to find the stable
region of EU with perfect fluid, so matter Lagrangian can be taken
as $\mathcal{L}_{m}=-p$ \cite{4}. In closed FRW universe background,
the field equations of this modified theory corresponding to matter
Lagrangian are obtained as
\begin{eqnarray}\nonumber
\frac{3}{a^{2}}(1+\dot{a}^{2})&=&\frac{1}{f_{R}+pf_{Q}}\left[\kappa^{2}
\rho+(\rho+p)f_{T}+3\left(\frac{a\ddot{a}+\dot{a}^{2}+1}{a^{2}}\right)
f_{R}+\frac{1}{2}\right.\\\nonumber&\times&\left.f(R,T,Q)-3\frac{\dot{a}}{a}
\partial_{t}f_{R}-\frac{3}{2}\left(\frac{4\dot{a}^{2}-a\ddot{a}+2}{a^{2}}
\right)\rho f_{Q} -\frac{3}{2}\right.\\\label{8}&\times&\left.\left
(\frac{a\ddot{a}+2\dot{a}^{2}}{a^{2}}\right)pf_{Q}
+\frac{3\dot{a}}{2a}\partial_{t}[(p-\rho)f_{Q}]\right],\\\nonumber
-2a\ddot{a}-(1+\dot{a}^{2})&=&\frac{1}{f_{R}+pf_{Q}}\left[\kappa^{2}a^{2}p
-\frac{a^{2}}{2}f(R,T,Q)-3(a\ddot{a}+\dot{a}^{2}+1)f_{R}\right.\\\nonumber&+&
\left.2a\dot{a}\partial_{t}f_{R}+a^{2}\partial_{tt}f_{R}+\frac{1}{2}\left
(\frac{4\dot{a}^{2}-a\ddot{a}+2}{a^{2}}\right)pf_{Q}+\frac{1}{2}(a\ddot{a}
\right.\\\label{9}&+&\left.2\dot{a}^{2})\rho f_{Q}+2a\dot{a}
\partial_{t}[(p +\rho)f_{Q}]+\frac{a^{2}}{2}\partial_{tt}[(\rho-p)f_{Q}]\right],
\end{eqnarray}
where dot shows derivative with respect to time. The conservation
equation (\ref{5}) with perfect fluid becomes
\begin{eqnarray}\nonumber
\dot{\rho}+3\frac{\dot{a}}{a}(\rho+p)&=&\frac{1}{2\kappa^{2}+3(f_{T}
-\frac{2\dot{a}^{2}+a\ddot{a}}{a^{2}}f_{Q})}\left[\left\{3\frac{\dot{a}}
{a^{3}}\left(a\ddot{a}-4\dot{a}^{2}\right)(\rho+p)\right.\right.
\\\nonumber&-&\left.\left.3(\frac{a\ddot{a}-\dot{a}^{2}}
{a^{2}})\partial_{t}p\right\}f_{Q}-2(\rho+p)\partial_{t}f_{T}+6(\frac{\dot{a}}{a})^{2}
(\rho+p)\right.\\\label{10}&\times&\left.\partial_{t}f_{Q}+\partial_{t}pf_{T}\right].
\end{eqnarray}

\section{Stability Analysis of Einstein Universe}

In this section, we consider linear homogeneous perturbations and
investigate the stability of EU in $f(R,T,Q)$ gravity. For this
purpose, we take $a(t)=a_{0}=constant$ for EU and the corresponding
field equations (\ref{8}) and (\ref{9}) reduce to
\begin{eqnarray}\nonumber
\frac{3}{a_{0}^{2}}&=&\frac{1}{f_{R}+p_{0}f_{Q}}\left[\kappa^{2}
\rho_{0}+(\rho_{0}+p_{0})f_{T}+\frac{1}{2}f(R_{0},T_{0},Q_{0})+\frac{3}
{a_{0}^{2}}f_{R}-\frac{3}{a_{0}^{2}}\rho_{0}f_{Q}\right],\\\label{11}\\\nonumber
-\frac{1}{a_{0}^{2}}&=&\frac{1}{f_{R}+p_{0}f_{Q}}\left[\kappa^{2}p_{0}
-\frac{1}{2}f(R_{0},T_{0},Q_{0})-\frac{3}{a_{0}^{2}}f_{R}+\partial_{tt}f_{R}
+\frac{1}{a_{0}^{2}}p_{0}f_{Q}\right.\\\label{12}
&+&\left.\frac{1}{2}\partial_{tt}[(\rho_{0}-p_{0})f_{Q}]\right],
\end{eqnarray}
where $R_{0}=R(a_{0})=-\frac{6}{a_{0}^{2}}$, $T_{0}=\rho_{0}-3p_{0}$
and $Q_{0}=6\frac{p_{0}}{a_{0}^{2}}$. Here $\rho_{0}$ and $p_{0}$
denote the unperturbed energy density and pressure, respectively. In
order to examine the stability regions, we consider linear EoS
defined as $p(t)=\omega\rho(t)$ ($\omega$ is the EoS parameter) and
introduce the expressions for linear perturbations in scale factor
and energy density depending only on time as follows
\begin{equation}\label{13}
a(t)=a_{0}+a_{0}\delta a(t),\quad
\rho(t)=\rho_{0}+\rho_{0}\delta\rho(t),
\end{equation}
where $\delta a(t)$ and $\delta\rho(t)$ express the perturbed scale
factor and energy density, respectively. We assume that $f(R,T,Q)$
is analytic and by applying Taylor series expansion for three
variables upto first order, this turns out to be
\begin{eqnarray}\nonumber
f(R,T,Q)&=&f(R_{0},T_{0},Q_{0})+f_{R}(R_{0},T_{0},Q_{0})\delta R
+f_{T}(R_{0},T_{0},Q_{0})\delta T\\\label{14}&+&f_{Q}
(R_{0},T_{0},Q_{0})\delta Q.
\end{eqnarray}
Using linear EoS, $\delta R$, $\delta T$ and $\delta Q$ become
\begin{equation}\label{15}
\delta R=-6(\delta\ddot{a}-2\frac{\delta a}{a_{0}^{2}}),\quad \delta
T=T_{0}\delta\rho_{0},\quad \delta Q=3\rho_{0}[(\omega
-1)\delta\ddot{a}-4\frac{\omega}{a_{0}^{2}}\delta a
+2\frac{\omega}{a_{0}^{2}}\delta\rho],
\end{equation}
where $\delta\ddot{a}=\frac{d^{2}}{dt^{2}}(\delta a)$. Substituting
Eqs.(\ref{11})-(\ref{15}) into the field equations (\ref{8}) and
(\ref{9}), we obtain the linearized perturbed equations as
\begin{eqnarray}\nonumber
&&6(f_{R}+\rho_{0}f_{Q})\delta a+a_{0}^{2}\rho_{0}[\kappa^{2}
+(1+\omega)f_{T}+\frac{1}{2}(1-3\omega)f_{T}-\frac{3}
{a_{0}^{2}}f_{Q}]\delta\rho=0,\\\label{16} \\\nonumber
&&2(f_{R}+\rho_{0}f_{Q})\delta\ddot{a}+\frac{2}{a_{0}^{2}}
(\rho_{0}\omega f_{Q}-f_{R})\delta a+\rho_{0}[\kappa^{2}\omega
-\frac{1}{2}(1-3\omega)f_{T}-\frac{\omega}{a_{0}^{2}}f_{Q}]
\delta\rho\\\label{17}&+&\frac{1}{2}\rho_{0}(1-\omega)f_{Q}
\delta\ddot{\rho}=0.
\end{eqnarray}
These express a direct relation between perturbed scale factor and
energy density.

In the following, we discuss the stability of EU for both conserved
and non-conserved EMT.

\subsection{Stability for Conserved EMT}

The conservation law does not hold in $f(R,T,Q)$ theory of gravity
like other modified theories having non-minimal coupling between
matter and geometry \cite{4,5}. We assume that this law holds in
this gravity for which the right hand side of Eq.(\ref{10}) becomes
zero and we obtain
\begin{eqnarray}\nonumber
&&\dot{p}f_{T}-2(\rho+p)\partial_{t}f_{T}+\frac{3}{a^{2}}\left[\frac{\dot{a}}
{a}(a\ddot{a}-4\dot{a}^{2})(\rho+p)-(a\ddot{a}-\dot{a}^{2})
\dot{p}\right]f_{Q}\\\label{18}&+&6(\frac{\dot{a}}{a})^{2}
(\rho+p)\partial_{t}f_{Q}=0.
\end{eqnarray}
From the standard conservation equation, we obtain the relation
defined by
\begin{equation}\label{19}
\delta\dot{\rho}(t)=-3(1+\omega)\delta\dot{a}(t).
\end{equation}
In order to have the perturbed field equation in the form of
perturbed scale factor, we eliminate $\delta\rho$ from
Eqs.(\ref{16}) and (\ref{17}) and then substitute Eq.(\ref{19}) in
the resulting equation, it follows that
\begin{eqnarray}\nonumber
&&\left[\frac{2}{a_{0}^{2}}(f_{R}-\omega\rho_{0}f_{Q})
\left[a_{0}^{2}\rho_{0}\left\{\kappa^{2}+(1+\omega)f_{T}
+\frac{1}{2}(1-3\omega)f_{T}-\frac{3}{a_{0}^{2}}
f_{Q}\right\}\right]+6\right.\\\nonumber&\times&
\left.\rho_{0}(f_{R}+\rho_{0}f_{Q})\left\{\kappa^{2}\omega
-\frac{1}{2}(1-3\omega)f_{T}-\frac{\omega}
{a_{0}^{2}}f_{Q}\right\}\right]\delta a+\left[a_{0}^{2}
\rho_{0}\left\{\kappa^{2}+(1+\omega)\right.\right.
\\\nonumber&\times&\left.\left.f_{T}+\frac{1}{2}
(1-3\omega)f_{T}-\frac{3}{a_{0}^{2}}f_{Q}\right\}\left\{\frac{3}{2}
\rho_{0}(1-\omega^{2})f_{Q}-2(f_{R}+\rho_{0}f_{Q})\right\}\right]
\delta\ddot{a}=0.\\\label{20}
\end{eqnarray}
To determine the expression for $a_{0}^{2}$, adding Eqs.(\ref{11})
and (\ref{12}) which leads to
\begin{equation}\label{21}
a_{0}^{2}=\frac{2f_{R}+\rho_{0}(3+\omega)f_{Q}}{\rho_{0}(1+\omega)
(\kappa^{2}+f_{T})}.
\end{equation}
Using this value in Eq.(\ref{20}), the perturbed field equation
takes the form
\begin{eqnarray}\nonumber
&&\left[2(f_{R}-\omega\rho_{0}f_{Q})\left[\rho_{0}\left\{\kappa^{2}
+(1+\omega)f_{T}-\left(\frac{3\rho_{0}(1+\omega)(\kappa^{2}+f_{T})}
{2f_{R}+\rho_{0}(3+\omega)f_{Q}}\right)f_{Q}\right.\right.\right.
\\\nonumber&+&\left.\left.\left.\frac{1}{2}(1-3\omega)f_{T}\right\}\right]+6\rho_{0}
(f_{R}+\rho_{0}f_{Q})\left\{\kappa^{2}\omega-\left(\frac{\omega\rho_{0}(1
+\omega)(\kappa^{2}+f_{T})}{2f_{R}+\rho_{0}(3+\omega)f_{Q}}\right)
\right.\right.\\\nonumber&\times&\left.\left.f_{Q}
-\frac{1}{2}(1-3\omega)f_{T}\right\}\right] \delta
a+\left[\rho_{0}\left(\frac{2f_{R}+\rho_{0}(3+\omega)
f_{Q}}{\rho_{0}(1+\omega)(\kappa^{2}+f_{T})}\right)\left\{\kappa^{2}
+(1+\omega)\right.\right.\\\nonumber&\times&\left.\left.
f_{T}+\frac{1}{2} (1-3\omega)f_{T}-\left(\frac{3\rho_{0}(1+\omega)
(\kappa^{2}+f_{T})}{2f_{R}+\rho_{0}(3+\omega)f_{Q}}\right)f_{Q}\right\}
\left\{\frac{3}{2}\rho_{0}(1-\omega^{2})f_{Q}\right.\right.
\\\label{22}&-&\left.\left.2(f_{R}+\rho_{0}f_{Q})\right\}\right]
\delta\ddot{a}=0.
\end{eqnarray}

As in other modified theories, we also have fourth-order perturbed
field equations in this theory. However, it vanishes due to the
presence of $Q$ term (product of Ricci tensor and EMT) as we are
assuming only the first-order linear terms. Thus we obtain a
second-order perturbation equation about $a(t)$ in this modified
theory. In the GR limit, i.e., for $f_{R}=1$ and $f_{T}=0=f_{Q}$,
Eq.(\ref{22}) reduces to the desired form given by
\begin{equation}\nonumber
2\delta\ddot{a}-\rho_{0}(1+\omega)(1+3\omega)\delta a=0.
\end{equation}
The solution of Eq.(\ref{22}) is helpful to examine the stability
modes of EU but due to a complicated nature of this theory, it would
be a difficult task. For this purpose, we consider a specific form
of $f(R,T,Q)$ gravity defined as follows \cite{7}
\begin{equation}\label{23}
f(R,T,Q)=R+f(T)+g(Q),
\end{equation}
where $f(T)$ and $g(Q)$ are the generic functions of $T$ and $Q$,
respectively. We assume that the conservation law holds for this
model. Consequently, the resulting second-order differential
equation is obtained using this particular form in Eq.(\ref{18}) as
\begin{equation}\nonumber
\omega f'(T)-2(1+\omega)Tf''(T)=0,
\end{equation}
where prime shows derivative with respect to $x(x=R$, or $T$, or
$Q)$. The solution of this equation is
\begin{equation}\label{24}
f(T)=\frac{c_{1}T(1+\omega)(2T(1+\omega))^{\frac{\omega}{2(1
+\omega)}}}{2+3\omega}+c_{2},
\end{equation}
where $c_{1}$ and $c_{2}$ are integration constants.

It is mentioned here that the conservation law holds only for this
unique expression of $f(T)$ in the model (\ref{23}). Now
substituting the values from Eqs.(\ref{23}) and (\ref{24}) in
(\ref{22}), the resulting differential equation takes the form
\begin{eqnarray}\nonumber
&&\left[6\{\Delta_{1}+\Delta_{2}\Delta_{3}+2\Delta_{3}^{2}
\Delta_{4}\}-3g'(Q)\{2\Delta_{5}+\Delta_{3}\Delta_{6}
-\Delta_{3}^{2}\Delta_{7}\}\right]\delta
a\\\label{25}&-&[(\Delta_{8}
+\Delta_{9})g'(Q)+2\Delta_{3}\Delta_{10}]\delta\ddot{a}=0,
\end{eqnarray}
where $\Delta_{i}$'s $(i=1,2,3,...,10)$ are
\begin{eqnarray}\nonumber
\Delta_{1}&=&\frac{\rho_{0}\kappa^{4}}{3}(1+4\omega
+3\omega^{2}),\\\nonumber \Delta_{2}&=&\rho_{0}\kappa^{2}
\left(1+\omega(10+7\omega)+2\rho_{0}\omega(1+3\omega
+\omega^{2})\right),\\\nonumber
\Delta_{3}&=&\frac{c_{1}}{2}\left(2\rho_{0}(1+\omega)(1
-3\omega)\right)^{\frac{\omega}{2(1+\omega)}},\\\nonumber
\Delta_{4}&=&\rho_{0}\omega(2+\omega+\rho_{0}(1+\omega
+\omega^{2})),\\\nonumber
\Delta_{5}&=&\rho_{0}^{2}\omega\kappa^{4}(1
+4\omega+3\omega^{2}),\\\nonumber
\Delta_{6}&=&\rho_{0}^{2}\kappa^{2}(8\omega(1+\omega)
+3(1-\omega^{4})),\\\nonumber
\Delta_{7}&=&\rho_{0}^{2}(\kappa^{2}\omega^{2}(1+3\omega)
-3(1+2\omega)+3\omega^{3}(2+\omega)),\\\nonumber
\Delta_{8}&=&\kappa^{2}\left(2(3+\omega)+\rho_{0}(1-4\omega
+3\omega^{2})\right),\\\nonumber
\Delta_{9}&=&9-\omega^{2}+\frac{\rho_{0}}{2}(9-13\omega
+7\omega^{2}-3\omega^{3}),\\\nonumber
\Delta_{10}&=&\frac{2\kappa^{2}}{\Delta_{3}}+(3-\omega).
\end{eqnarray}
The solution of Eq.(\ref{25}) is given by
\begin{equation}\nonumber
\delta a(t)=b_{1}e^{\Omega t}+b_{2}e^{-\Omega t}.
\end{equation}
Here $b_{1}$ and $b_{2}$ are constants of integration and the
parameter $\Omega$ represents the frequency of small perturbation
which is of the form
\begin{equation}\label{26}
\Omega^{2}=\frac{6(\Delta_{1}+\Delta_{2}\Delta_{3}+2\Delta_{3}^{2}
\Delta_{4})-3g'(Q)(2\Delta_{5}+\Delta_{3}\Delta_{6}-\Delta_{3}^{2}
\Delta_{7})}{(\Delta_{8}+\Delta_{9})g'(Q)+2\Delta_{3}\Delta_{10}}.
\end{equation}
In order to avoid the exponential increase in $\delta a(t)$ or
collapse, the parameter $\Omega^{2}<0$, which leads to the stability
of EU. In general relativistic limit, this frequency is given by
\begin{equation}\nonumber
\Omega^{2}=\frac{1}{2}\kappa^{2}\rho_{0}(1+\omega)(1+3\omega),
\end{equation}
which shows the stable solution in the range
$-1<\omega<-\frac{1}{3}$ \cite{17}.
\begin{figure}\center
\epsfig{file=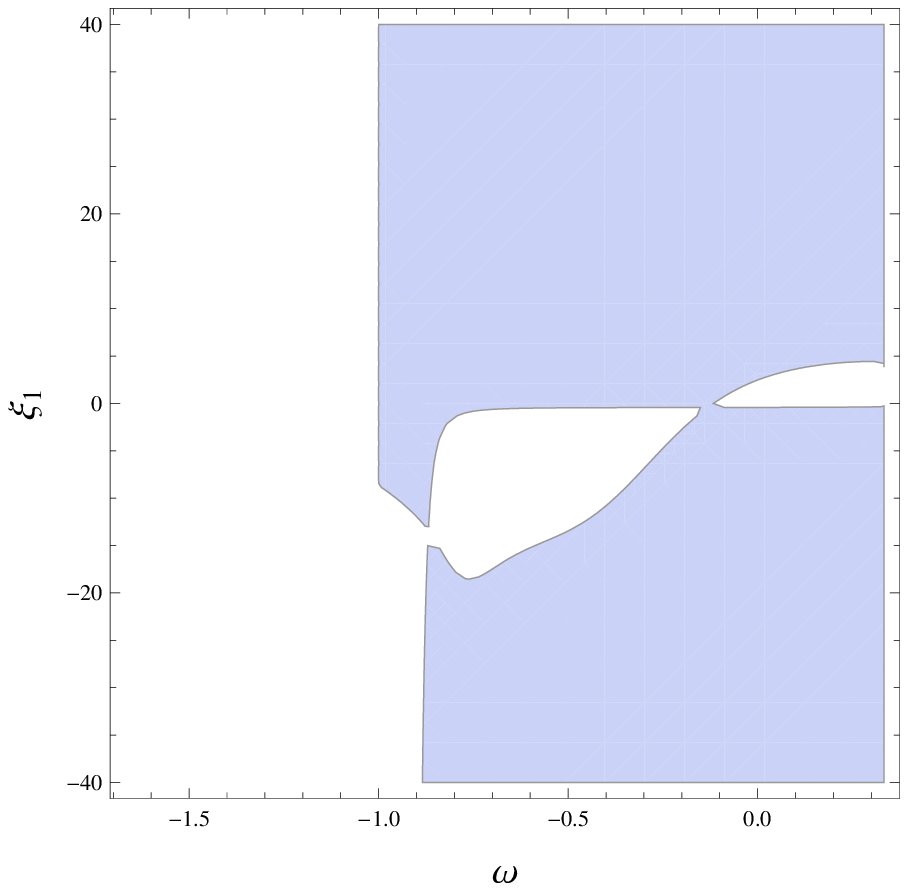,width=0.47\linewidth}\epsfig{file=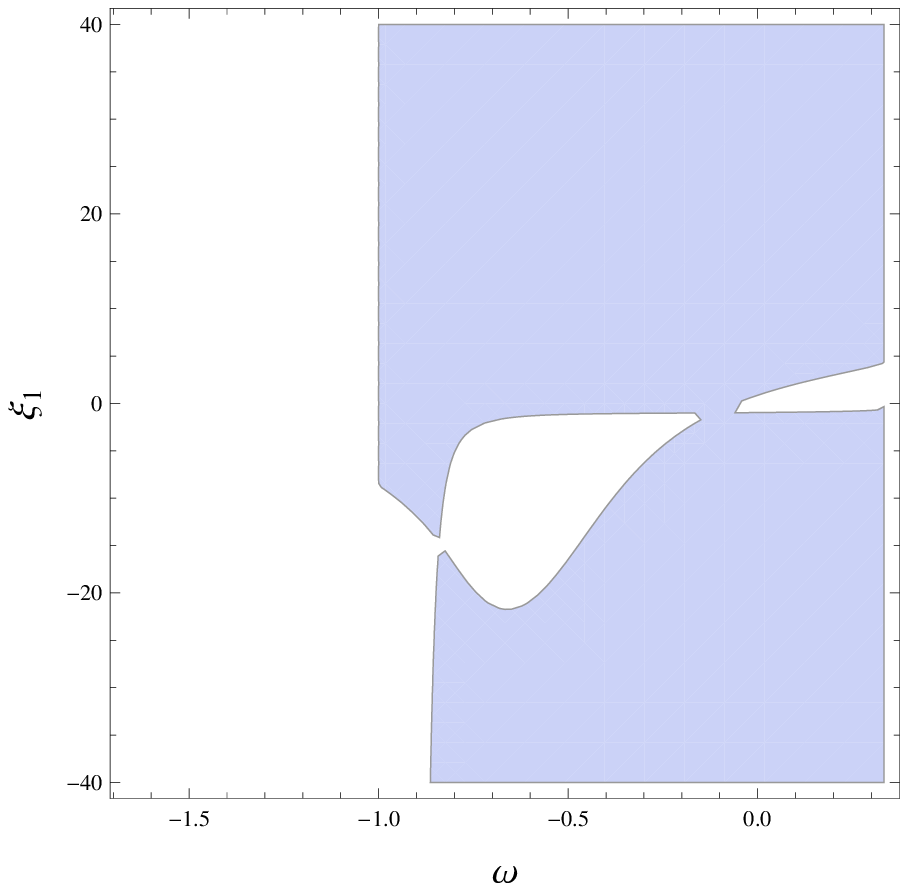,width=0.47\linewidth}
\epsfig{file=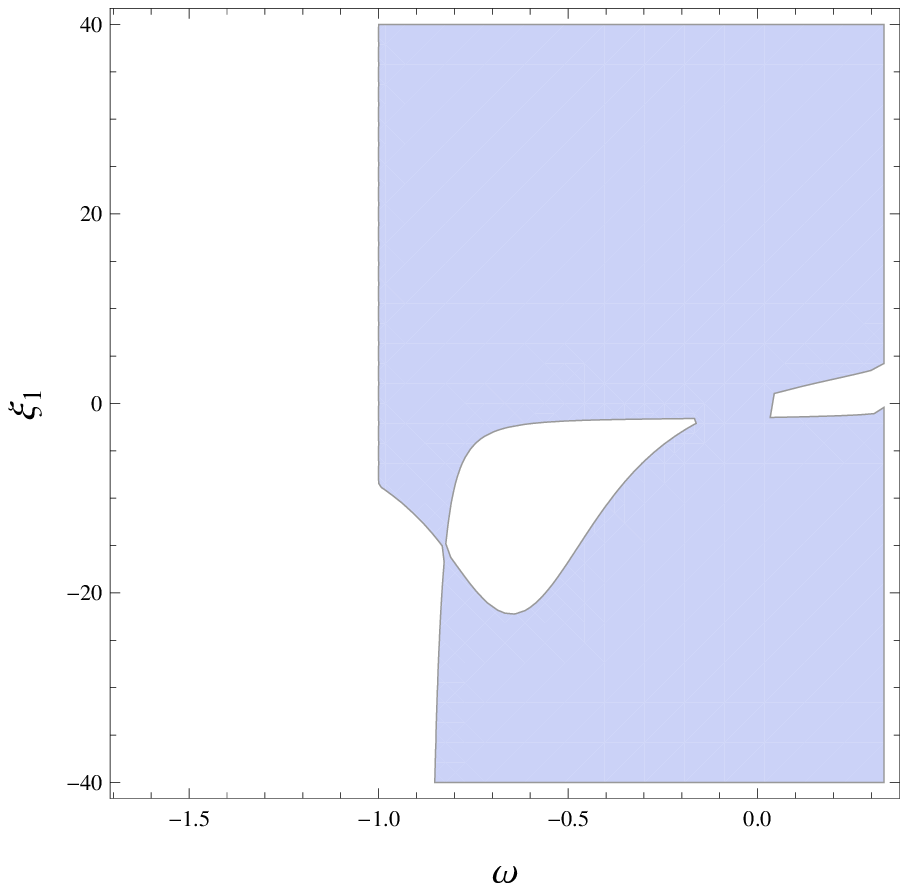,width=0.47\linewidth}\\
\caption{Plots of stable regions in $(\omega,\xi_{1})$ space for
$\Omega^{2}$ with $c_{1}=1$ (left), $c_{1}=4$ (right) and $c_{1}=7$
(below).}
\end{figure}
\begin{figure}\center
\epsfig{file=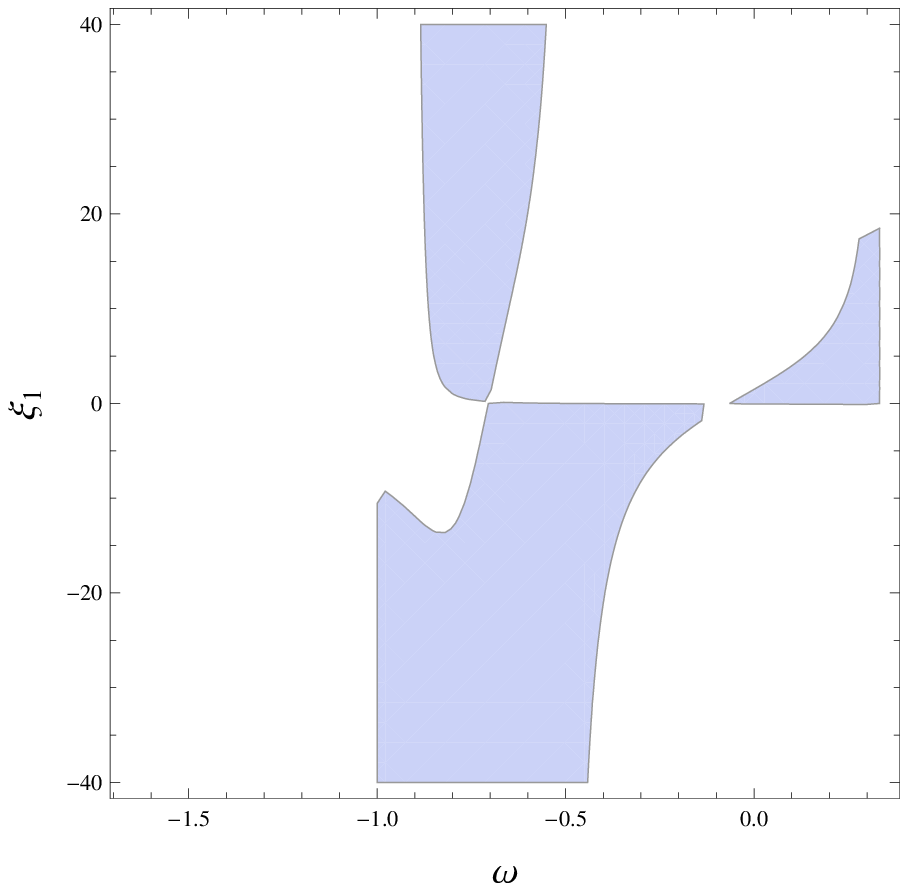,width=0.47\linewidth}\epsfig{file=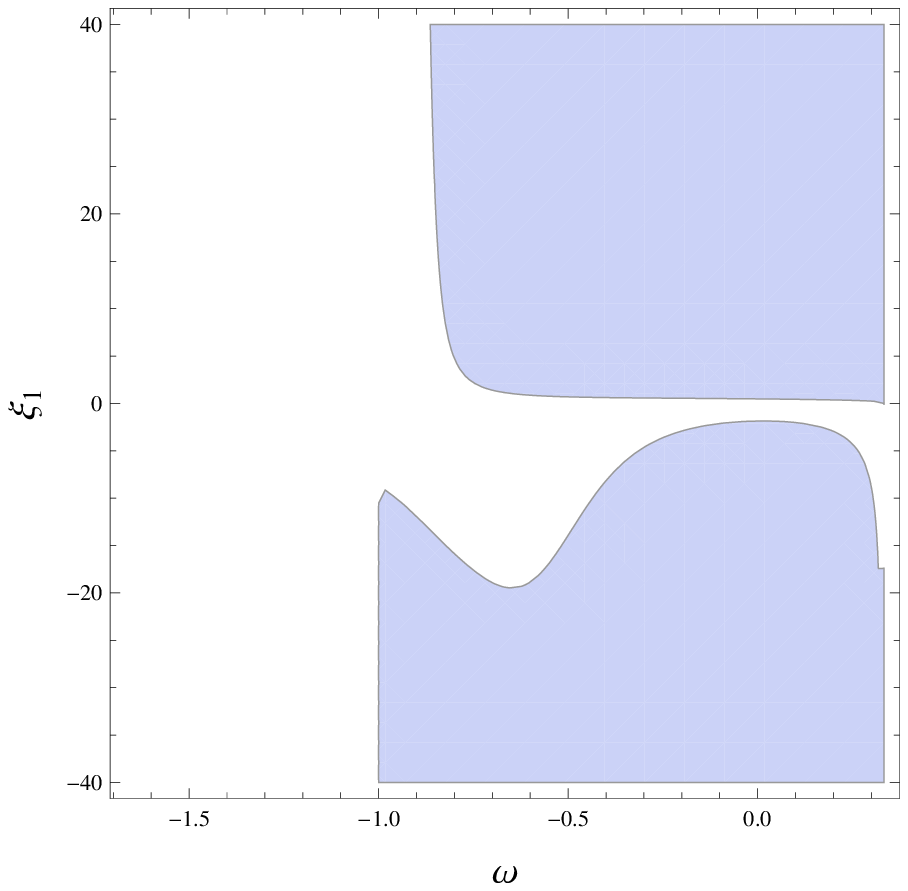,width=0.47\linewidth}
\epsfig{file=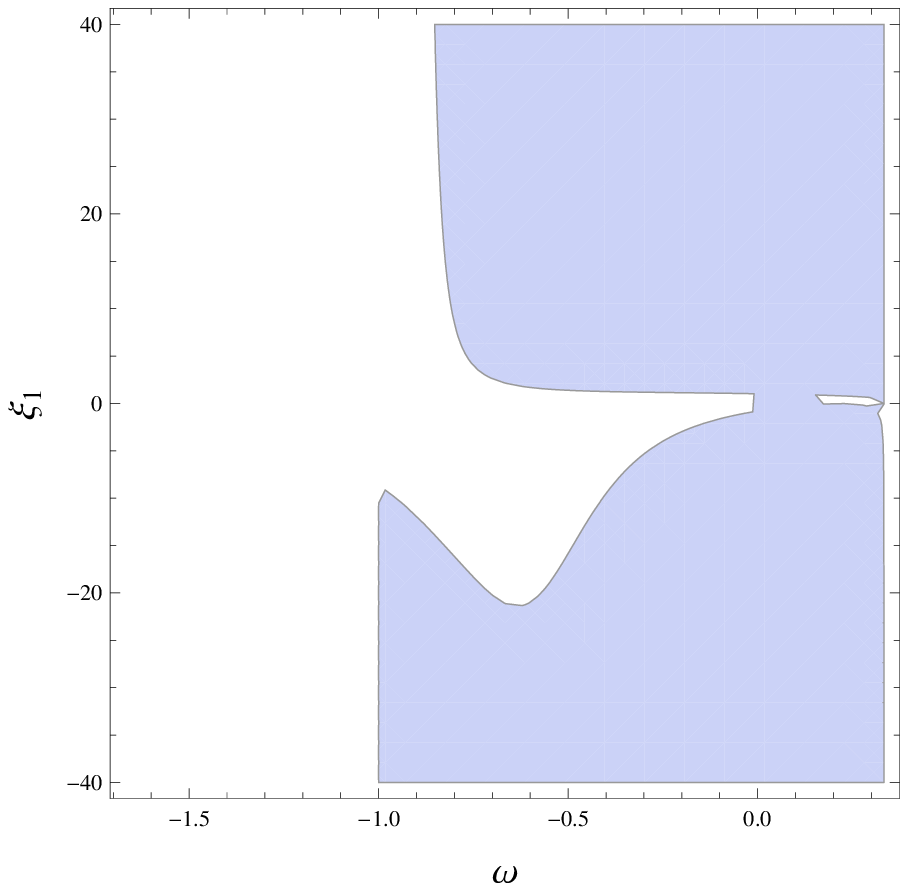,width=0.47\linewidth}\\
\caption{Plots of stable regions in $(\omega,\xi_{1})$ space for
$\Omega^{2}$ with $c_{1}=-1$ (left), $c_{1}=-4$ (right) and
$c_{1}=-7$ (below).}
\end{figure}

To analyze the graphical behavior of stable modes of EU, we take
$\kappa^{2}=1$, $\rho_{0}=0.3$ \cite{30} and $g'(Q)=\xi_{1}$ as a
new parameter. Figure \textbf{1} shows the existence of stable EU
for $\Omega^{2}$ under linear perturbations corresponding to
different positive values of $c_{1}$. It is observed that the stable
EU exists for all values of $\omega\geq-1$ and these stability
regions are becoming more smooth with increasing value of
integration constant $c_{1}$. The graphical behavior in Figure
\textbf{2} describes the EU and its stability for negative values of
$c_{1}$. It is found that for $c_{1}=-1$, the stable modes exist in
the range of $-1<\omega<\frac{1}{3}$ and with decreasing value of
$c_{1}$, the graphs show more stable regions towards positive values
of EoS parameter.

\subsection{Stability for Non-Conserved EMT}

In this section, we discuss the stability when EMT is not conserved.
Here we consider another specific model which consists of linear
form of $R$ and generic function $f(Q)$ defined as follows \cite{7}
\begin{equation}\label{27}
f(R,T,Q)=\alpha R+f(Q),
\end{equation}
where $\alpha$ is an arbitrary constant. For this model, the
perturbed field equations (\ref{16}) and (\ref{17}) lead to
\begin{eqnarray}\label{28}
&&6\left[\alpha+\rho_{0}f'(Q)\right]\delta
a+a_{0}^{2}\rho_{0}\left[\kappa^{2}
-\frac{3}{a_{0}^{2}}f'(Q)\right]\delta\rho=0,\\\nonumber
&&\left[2\alpha+\rho_{0}(\frac{1+3\omega^{2}}{2})f'(Q)\right]\delta\ddot{a}
+\frac{2}{a_{0}^{2}}\left[\rho_{0}\omega f'(Q)-\alpha\right]\delta a
\\\label{29}&+&\rho_{0}\omega[\kappa^{2}
-\frac{1}{a_{0}^{2}}f'(Q)]\delta\rho=0.
\end{eqnarray}
These equations represent the relationship between perturbed scale
factor and energy density perturbations. The differential equation
in perturbed scale factor is obtained by eliminating $\delta\rho$
from Eqs.(\ref{28}) and (\ref{29}) as
\begin{eqnarray}\nonumber
&&a_{0}^{2}\left[2\alpha\kappa^{2}+\left\{\rho_{0}\kappa^{2}(\frac{1+3
\omega^{2}}{2})-\frac{6\alpha}{a_{0}^{2}}\right\}f'(Q)\right]\delta\ddot{a}
+2\left[\left\{\frac{3\alpha}{a_{0}^{2}}(1+\omega)\right.
\right.\\\label{30}&-&\left.\left.2\rho_{0}\omega\kappa^{2}\right\}f'(Q)
-\alpha\kappa^{2}(1+3\omega)\right]\delta a=0.
\end{eqnarray}
The addition of static field Eqs.(\ref{11}) and (\ref{12})
corresponding to (\ref{27}) yields
\begin{equation}\label{31}
a_{0}^{2}=\frac{2\alpha+\rho_{0}(3+\omega)f'(Q)}{\rho_{0}\kappa^{2}(1+\omega)}.
\end{equation}
Inserting the value of $a_{0}^{2}$ in Eq.(\ref{30}), the resulting
differential equation is
\begin{eqnarray}\nonumber
&&2\kappa^{2}\rho_{0}(1+\omega)\left[\alpha(1+3\omega)+4\omega\rho_{0}f'(Q)
\right]\delta a-\left[4\alpha^{2}+\alpha\rho_{0}(3\omega^{2}-2\omega
\right.\\\label{32}&+&\left.7)f'(Q)\right]\delta\ddot{a}=0,
\end{eqnarray}
whose solution is obtained as
\begin{equation}\label{33}
\delta a(t)=a_{1}e^{\bar{\Omega} t}+a_{2}e^{-\bar{\Omega} t},
\end{equation}
where $a_{j}$'s $(j=1,2)$ are integration constants and frequency of
small perturbation $(\bar{\Omega})$ is of the form
\begin{equation}\nonumber
\bar{\Omega}^{2}=\frac{2\kappa^{2}\rho_{0}(1+\omega)\left(\alpha(1
+3\omega)+4\omega\xi_{2}\right)}{4\alpha^{2}+\alpha(3\omega^{2}-2\omega
+7)\xi_{2}},
\end{equation}
where $\xi_{2}=\rho_{0}f'(Q)$.
\begin{figure}\center
\epsfig{file=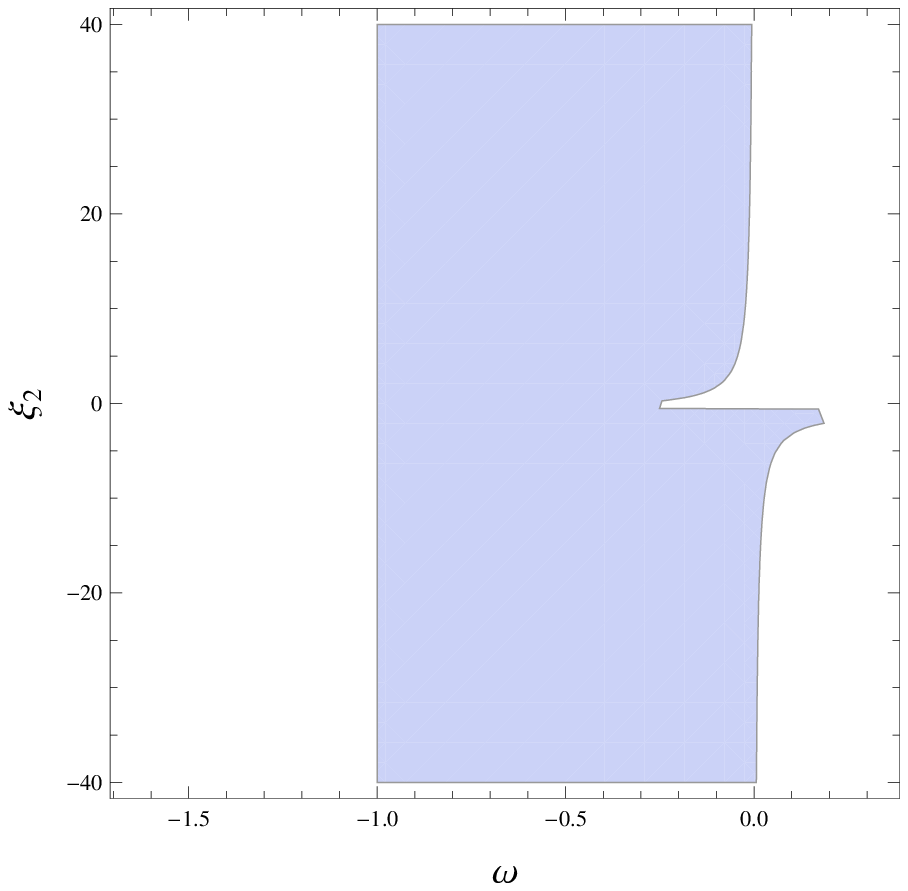,width=0.47\linewidth}\epsfig{file=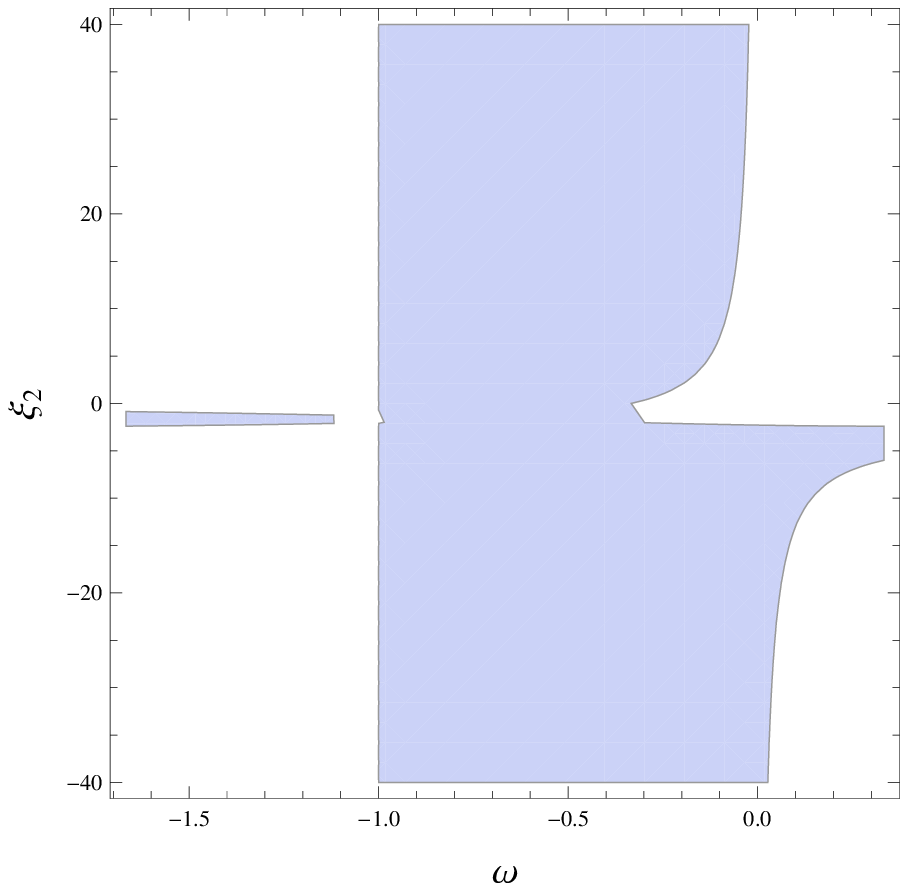,width=0.47\linewidth}
\epsfig{file=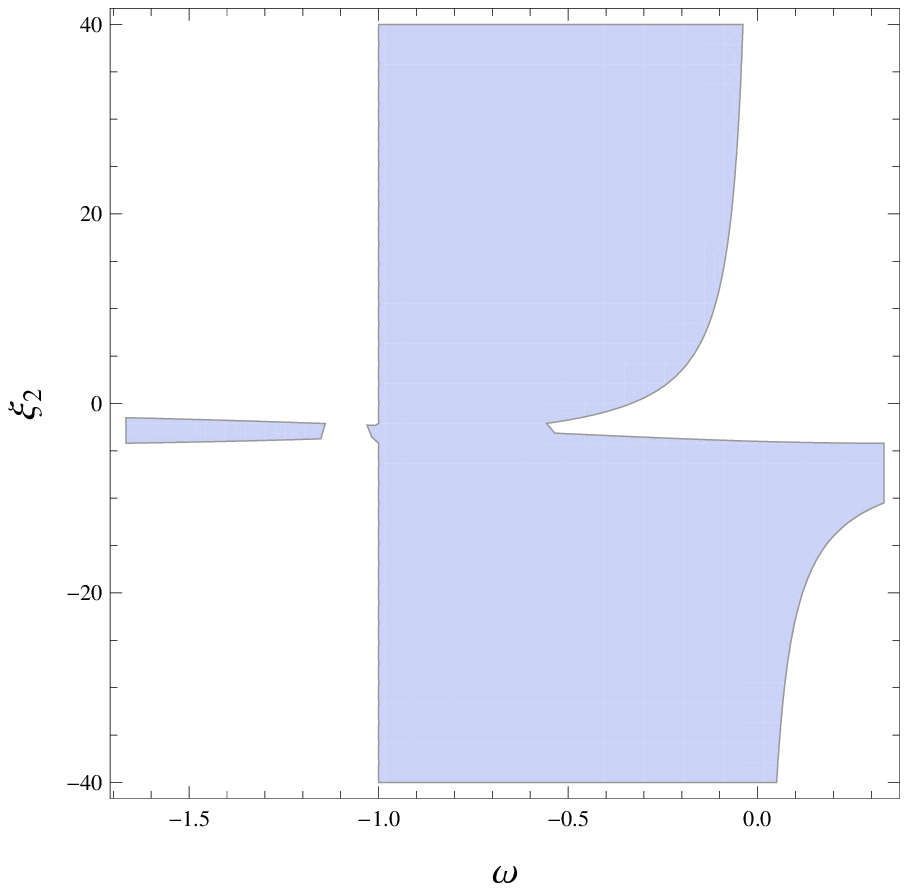,width=0.47\linewidth}\\
\caption{Plots of stable regions in $(\omega,\xi_{2})$ space for
$\bar{\Omega}^{2}$ with $\alpha=1$ (left), $\alpha=4$ (right) and
$\alpha=7$ (below).}
\end{figure}
\begin{figure}\center
\epsfig{file=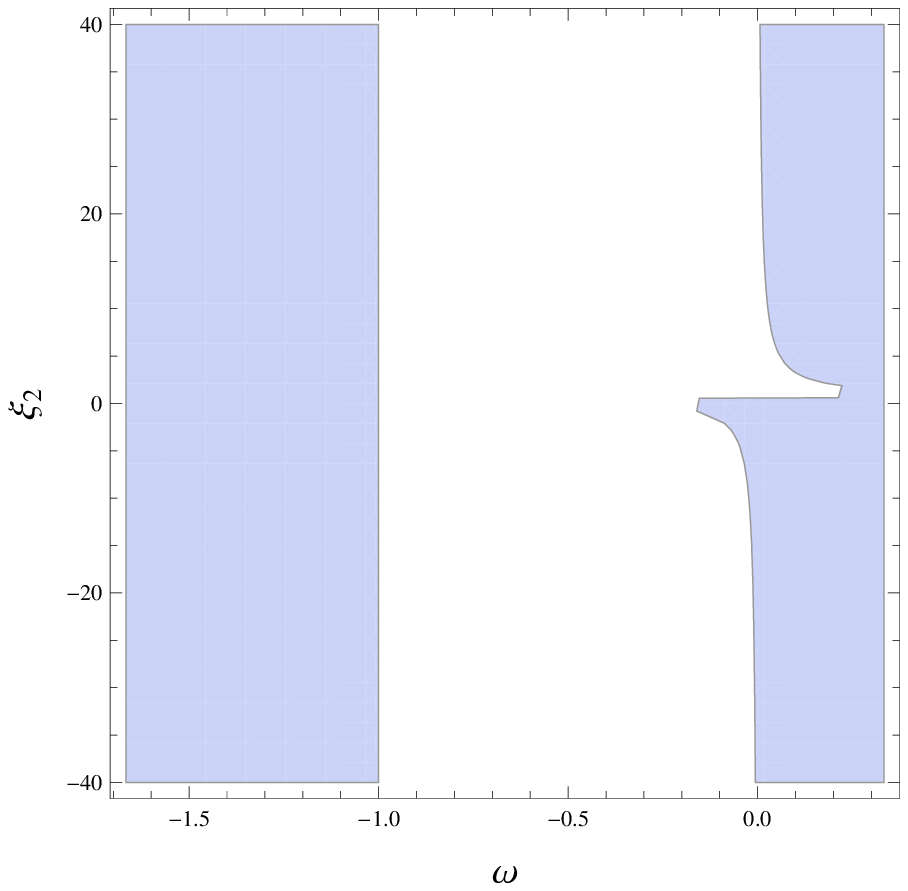,width=0.47\linewidth}\epsfig{file=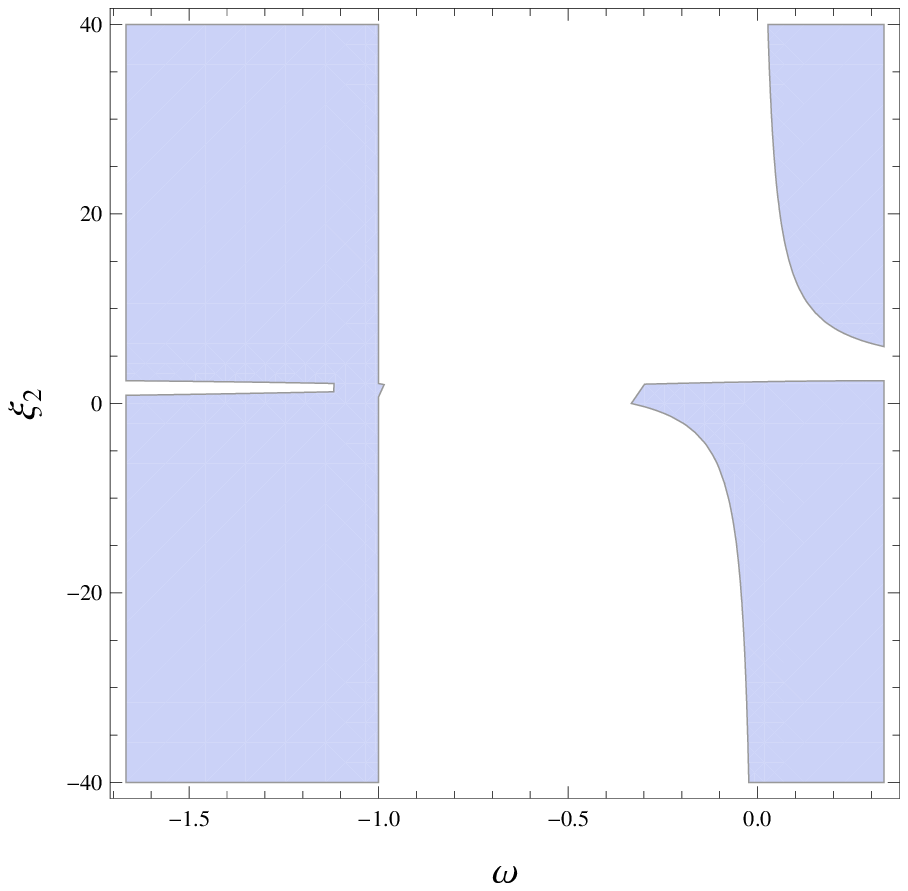,width=0.47\linewidth}
\epsfig{file=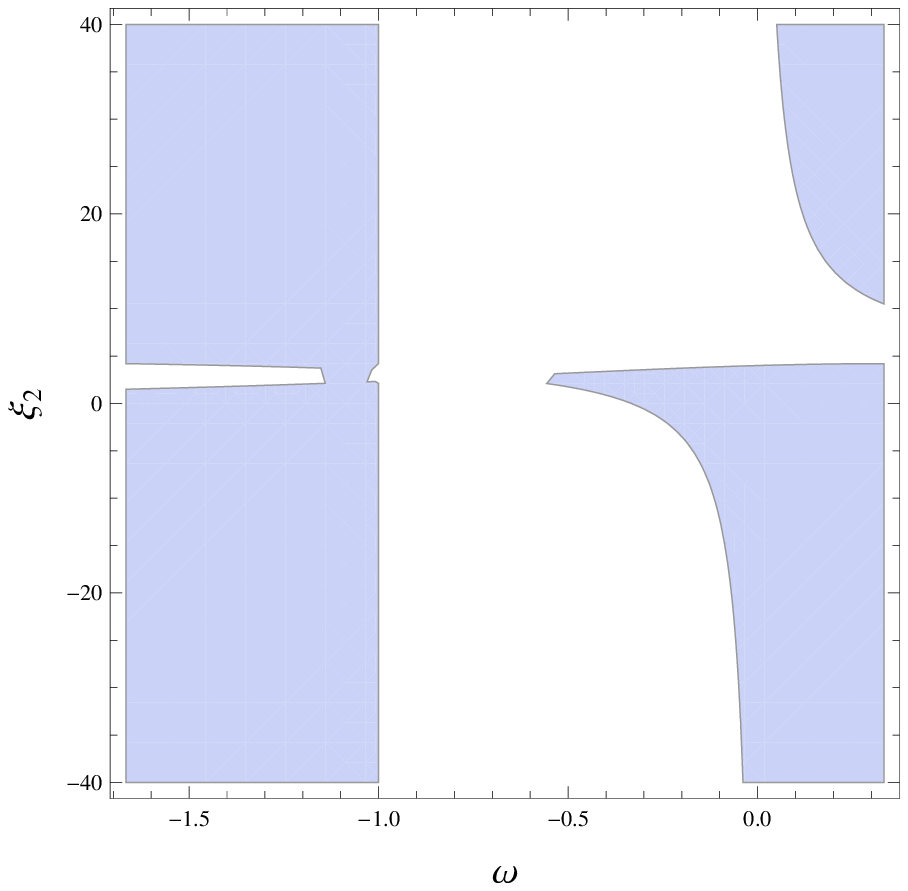,width=0.47\linewidth}\\
\caption{Plots of stable regions in $(\omega,\xi_{2})$ space for
$\bar{\Omega}^{2}$ with $\alpha=-1$ (left), $\alpha=-4$ (right) and
$\alpha=-7$ (below).}
\end{figure}

When $\alpha=1$ and $f(Q)=0$, the frequency $\bar{\Omega}^{2}$
regains the result of GR as calculated in the conserved case. The
graphical interpretation of stable regions against homogeneous
perturbations for different values of arbitrary constant is given in
Figures \textbf{3} and \textbf{4}. Figure \textbf{3} shows that for
$\alpha=1$, the stable regions appear only for negative values of
$\omega$ but for greater values of $\alpha$, we obtain some stable
region also for positive values of $\omega$. Figure \textbf{4}
indicates the existence of stable modes of EU for negative values of
arbitrary constant and the stability increases with decreasing value
of $\alpha$ for both positive as well as negative values of
$\omega$.

\section{Concluding Remarks}

In this paper, we have studied the stability of EU with closed FRW
universe model and perfect fluid in the framework of $f(R,T,Q)$
gravity. We have formulated static and perturbed field equations
using scalar homogeneous perturbations about energy density and
scale factor. These equations are parameterized by linear EoS
parameter. The second order perturbed differential equations are
constructed whose solutions provide the existence and stability
regions of EU for particular $f(R,T,Q)$ models. We have analyzed
both conserved as well as non-conserved EMT cases against
perturbations scheme. We have found stable results for some positive
as well as negative values of model parameters.

We conclude that stable modes of EU exist against homogeneous scalar
perturbations for all values of EoS parameter if the model
constraints are chosen appropriately in $f(R,T,Q)$ gravity. The
stable solutions of EU against vector perturbations are also exist
for all equations of state because any initial vector perturbations
remain frozen. It is worthwhile to mention here that the range of
EoS parameter is greatly enhanced as compared to that in $f(R)$
gravity and more stable regions are found as compared to $f(R,T)$
theory due to the presence of generic function of $Q$. Like all
other modified theories, our results also reduce to GR in the
absence of dark source terms. It would be interesting to extend our
results with the inhomogeneous perturbations around EU which indeed
could provide a richer structure for stability analysis of Einstein
cosmos in $f(R,T,Q)$ gravity.

\end{document}